\documentclass[aps,preprint]{revtex4}%
\usepackage{amssymb}
\usepackage{amsfonts}
\usepackage{amsmath}
\usepackage{graphicx}%
\providecommand{\U}[1]{\protect\rule{.1in}{.1in}}
\newtheorem{theorem}{Theorem}
\newtheorem{acknowledgement}[theorem]{Acknowledgement}

\begin{document}
\title{Chern--Simons--Antoniadis--Savvidy forms and standard supergravity}
\author{F. Izaurieta}
\email{fizaurie@udec.cl}
\author{P. Salgado}
\email{pasalgad@udec.cl}
\author{S. Salgado}
\email{sesalgado@udec.cl}
\affiliation{Departamento de F\'{\i}sica, Universidad de Concepci\'{o}n, Casilla 160-C,
Concepci\'{o}n, Chile}

\begin{abstract}
In the context of the so called the Chern--Simons--Antoniadis--Savvidy (ChSAS)
forms, we use the methods for FDA decomposition in 1-forms to construct a
four-dimensional ChSAS supergravity action for the Maxwell superalgebra. On
the another hand, we use the Extended Cartan Homotopy Formula to find a method
that allows the separation of the ChSAS action into bulk and boundary
contributions and permits the splitting of the bulk Lagrangian into pieces
that reflect the particular subspace structure of the gauge algebra.

\end{abstract}
\maketitle

\section{Introduction}

In Refs.~\cite{savv6,savv7,savv8,savv9} Antoniadis, Konitopoulos and Savvidy
introduced a procedure to construct gauge invariant, background-free gauge
forms. \ The integrals of these forms over the corresponding space-time
coordinates provides new topological actions \ that we have called
Chern--Simons--Antoniadis--Savvidy (ChSAS) actions, which generalize the usual
Chern--Simons theory. Of special interest are those which can be constructed
in even dimensions.

Using calculation methods for Free Differential Algebra (FDA) that allow the
decomposition of $p$-forms in 1-forms developed in Ref. \cite{dauria} and
applied in Refs. \cite{azcar,salg,ims}, we construct a four-dimensional ChSAS
supergravity action for Maxwell superalgebra. \ 

Following Ref. \cite{irs}\ it is found, in the context of the so called the
ChSAS theory,\ \ a subspace separation method for the Lagrangian. The method
is based on the iterative use of the generalized Extended Cartan Homotopy
Formula, and allows one to separate the action in bulk and boundary
contributions, and systematically split the Lagrangian in appropriate
reflection of the subspace structure of the gauge algebra. In order to apply
the method, one must regard ChSAS forms as a particular case of more general
objects known as generalized transgression forms.

This work is organized as follows: in Section 2, we briefly review the
principal aspects of transgression and Chern--Simons--Antoniadis--Savvidy
forms. In section 3, we use the calculation methods for Free Differential
Algebra (FDA) that allow the decomposition of $p$-forms in 1-forms developed
in Ref. \cite{dauria} and applied in Refs. \cite{azcar,salg,ims} to construct
a four-dimensional Chern--Simons--Antoniadis--Savvidy supergravity action for
Maxwell superalgebra. Section 4 presents the generalized Extended Cartan
Homotopy Formula and shows how a subspace separation method that allows for a
deeper understanding of the ChSAS Lagrangian can be built upon it. We finish
in Section~5 with conclusions and some considerations on future possible developments.

\section{\textbf{Chern--Simons--Antoniadis--Savvidy forms }in $\left(
2n+2\right)  $-dimensions}

The idea of extending the Yang--Mills fields to higher rank tensor gauge
fields was used in Ref.~\cite{savv6} in order to construct gauge invariant and
metric independent forms in higher dimensions. These forms are analogous to
the Pontryagin--Chern forms in Yang--Mills gauge theory. These results were
generalized in Refs.~\cite{savv7,savv8,savv9}, where the authors found closed
invariant forms similar to the Pontryagin--Chern forms in non-abelian tensor
gauge field theory. These forms are based on non-abelian tensor gauge fields
and are polynomials on the corresponding curvature forms.

A Lie algebra valued $1$-form connection $A$ can be written making more or
less explicit the dependence on the Lie algebra generator basis $T_{a}$ or the
basis of 1-forms $\mathrm{d}x^{\mu}$,%
\[
A=A_{\mu}\otimes\mathrm{d}x^{\mu}=A^{a}{}_{\mu}T_{a}\otimes\mathrm{d}x^{\mu}.
\]
The same is true for the $2$-form $B$ gauge potential $B=\frac{1}{2}B_{\mu\nu
}\otimes\mathrm{d}x^{\mu}\mathrm{d}x^{\nu}=\frac{1}{2}B^{a}{}_{\mu\nu}%
T_{a}\otimes\mathrm{d}x^{\mu}\mathrm{d}x^{\nu}$. The corresponding 2-form and
3-form \textquotedblleft curvatures\textquotedblright\ are given by
$F=\frac{1}{2!}F_{\mu\nu}\otimes\mathrm{d}x^{\mu}\mathrm{d}x^{\nu}$ and
$H=\frac{1}{3!}H_{\mu\nu\lambda}\otimes\mathrm{d}x^{\mu}\mathrm{d}x^{\nu
}\mathrm{d}x^{\lambda}$ respectively, where%
\begin{equation}
F=\mathrm{d}A+A^{2},\quad H=\mathrm{D}B=\mathrm{d}B+[A,B]. \label{def F y H}%
\end{equation}
The curvatures $F$ and $H$ satisfy the Bianchi identities%
\begin{equation}
\mathrm{D}F=0,\text{ \ \ }\mathrm{D}H+[B,F]=0. \label{id Bianchi}%
\end{equation}
The infinitesimal, non-abelian gauge transformations for the generalized gauge
fields are given by%
\begin{equation}
\delta A=\mathrm{D}\xi_{0},\text{ \ \ }\delta B=\mathrm{D}\xi_{1}+[B,\xi_{0}],
\label{transf A}%
\end{equation}
where $\xi_{0}$ and $\xi_{1}$ are a $0$-form and a $1$-form gauge parameters
respectively \cite{savv6}. Under these gauge transformations, the curvatures
transform as~\cite{savv7},%
\begin{align}
\delta F  &  =\mathrm{D}(\delta A)=\left[  F,\xi_{0}\right]  ,\label{one}\\
\delta H  &  =\mathrm{D}(\delta B)+\left[  \delta A,B\right]  . \label{two}%
\end{align}

It may be of interest to note that we have used the definition of the
conmutator for differential forms given by
\[
\left[  X,Y\right]  =XY-\left(  -1\right)  ^{pq}YX,
\]
where $X$ is a $p$-form and $Y$ is a $q$-form.

In Refs.~\cite{savv6,savv7} there were found closed invariant forms similar to
the Pontryagin--Chern forms in non-abelian tensor gauge field theory. \ In
particular, it was found that there exists a gauge invariant
metric-independent invariant in $\left(  2n+3\right)  $-dimensional space-time%
\begin{equation}
\Gamma_{2n+3}=\langle F^{n}H\rangle. \label{drei}%
\end{equation}

\textbf{Chern--Weil theorem in the }$\left(  2n+2\right)  $%
\textbf{-dimensional case}: the theorem ingredients are: $(i)$ Two Lie-algebra
valued, $1$-forms connections $A_{0}$ and $A_{1}$. Their curvatures are given
by, $F_{0}=\mathrm{d}A_{0}+A_{0}^{2}$ and $F_{1}=\mathrm{d}A_{1}+A_{1}^{2}$,
respectively.$\ (ii)$ Two Lie-algebra valued, generalized $2$-forms gauge
fields $B_{0}$ and $B_{1}$. Their generalized curvatures are given by
$H_{0}=\mathrm{d}B_{0}+[A_{0},B_{0}]$ and $H_{1}=\mathrm{d}B_{1}+[A_{1}%
,B_{1}]$ respectively $(iii)$\ In terms of these fundamental ingredients, it
is possible to define the differences $\Theta=A_{1}-A_{0}\ $and $\Phi
=B_{1}-B_{0}$, and the interpolating connections $A_{t}=A_{0}+t\Theta$ and
$B_{t}=B_{0}+t\Phi$ with $0\leq t\leq1$\textbf{. }Their corresponding
curvatures are given by%
\begin{equation}
F_{t}=\mathrm{d}A_{t}+A_{t}^{2},\text{ \ \ }H_{t}=\mathrm{D}_{t}%
B_{t}=\mathrm{d}B_{t}+[A_{t},B_{t}],
\end{equation}
which satisfy the conditions%
\begin{equation}
\frac{\mathrm{d}}{\mathrm{d}t}F_{t}=\mathrm{D}_{t}\Theta,\text{ \ \ }%
\frac{\mathrm{d}}{\mathrm{d}t}H_{t}=\mathrm{D}_{t}\Phi+\left[  \Theta
,B_{t}\right]  . \label{31}%
\end{equation}

\textbf{Theorem \cite{ims}:\ }Let $A_{0}$ and $A_{1}$ be two gauge connection
$1$-forms, and let\textbf{\ }$F_{0}$\textbf{\ }and\textbf{ }$F_{1}$ be their
corresponding $2$-forms curvature. \ Let $B_{0}$ and $B_{1}$ be two gauge
connection 2-forms and let $H_{0}$ and $H_{1}$ be their corresponding
curvature $3$-forms. Then, the difference $\Gamma_{2n+3}^{(1)}-\Gamma
_{2n+3}^{(0)}$ is an exact form%
\begin{equation}
\Gamma_{2n+3}^{(1)}-\Gamma_{2n+3}^{(0)}=\langle F_{1}^{n}H_{1}\rangle-\langle
F_{0}^{n}H_{0}\rangle=\mathrm{d}\mathfrak{T}^{\left(  2n+2\right)  }%
(A_{0},B_{0};A_{1},B_{1}), \label{six}%
\end{equation}
where%
\begin{equation}
\mathfrak{T}^{\left(  2n+2\right)  }(A_{0},B_{0};A_{1},B_{1})=\int_{0}%
^{1}\mathrm{d}t\left(  n\langle F^{n-1}\Theta H_{t}\rangle+\langle F_{t}%
^{n}\Phi\rangle\right)  \label{sieben}%
\end{equation}
is what we call \textquotedblleft Antoniadis--Savvidy transgression
form\textquotedblright. A proof can be found in Ref. \cite{ims}. Following the
same procedure followed in the case of the Chern--Simons forms, we define the
$\left(  2n+2\right)  $-Chern--Simons--Antoniadis--Savvidy form as%
\begin{equation}
\mathfrak{C}_{\mathrm{ChSAS}}^{\left(  2n+2\right)  }=\mathfrak{T}^{\left(
2n+2\right)  }(A,B;0,0)=\int_{0}^{1}\mathrm{d}t\langle nAF_{t}^{n-1}%
H_{t}+BF_{t}^{n}\rangle. \label{CSS general}%
\end{equation}
This result agrees with the expression found by Antoniadis and Savvidy in
Refs.~\cite{savv6,savv7}. \ From eq.~\textbf{(}\ref{CSS general}\textbf{), }we
have for the $n=1$ case \cite{savv7},%
\begin{equation}
\mathfrak{C}_{\mathrm{ChSAS}}^{(4)}=\int_{0}^{1}\mathrm{d}t\langle
AH_{t}+F_{t}B\rangle=\langle FB\rangle-\mathrm{d}\left\langle AB\right\rangle
. \label{accion CS 4 formas}%
\end{equation}
It is interesting to notice that transgression forms (both, standard ones and
the above generalization) are defined globally on the spacetime basis manifold
of the principal bundle, and are off-shell gauge invariant. Chern--Simons
forms (both, standard ones and the Antoniadis--Savvidy generalization) are
locally defined and are off-shell gauge invariant only up to boundary terms
(i.e., quasi-invariants).

\section{\textbf{Chern--Simons--Antoniadis--Savvidy form for Maxwell
superalgebra}}

Now we will use this construction for the particular case of the Maxwell
superalgebra, in order to show the connection between
eq.~(\ref{accion CS 4 formas}) and supergravity in $D=4$.

\subsection{ $\mathit{sM}_{4}$ Maxwell superalgebra}

The minimal Maxwell superalgebra $\mathit{sM}_{4}$ in $D=4$ is an algebra
whose generators $\{P_{a},J_{ab},Z_{ab},\tilde{Z}_{ab},Q_{\alpha}%
,\Sigma_{\alpha}\}$ satisfy the following commutation relation \cite{eve1,azc}%
\begin{align*}
\left[  J_{ab},J_{cd}\right]   &  =\eta_{bc}J_{ad}+\eta_{ad}J_{bc}-\eta
_{bd}J_{ac}-\eta_{ac}J_{bd},\\
\left[  J_{ab},P_{c}\right]   &  =\eta_{bc}P_{a}-\eta_{ac}P_{b},\text{
\ \ }\left[  P_{a},P_{b}\right]  =Z_{ab},\\
\left[  J_{ab},Z_{cd}\right]   &  =\eta_{bc}Z_{ad}+\eta_{ad}Z_{bc}-\eta
_{bd}Z_{ac}-\eta_{ac}Z_{bd},\\
\left[  P_{a},Q_{\alpha}\right]   &  =-\frac{1}{2}\left(  \gamma_{a}%
\Sigma\right)  _{\alpha},\text{ \ \ }\left[  J_{ab},Q_{\alpha}\right]
=-\frac{1}{2}\left(  \gamma_{ab}Q\right)  _{\alpha},\\
\left[  J_{ab},\Sigma_{\alpha}\right]   &  =-\frac{1}{2}\left(  \gamma
_{ab}\Sigma\right)  _{\alpha},\text{ \ \ }\left[  \tilde{Z}_{ab},Q_{\alpha
}\right]  =-\frac{1}{2}\left(  \gamma_{ab}\Sigma\right)  _{\alpha},\\
\left\{  Q_{\alpha},Q_{\beta}\right\}   &  =-\frac{1}{2}\left[  \left(
\gamma^{ab}C\right)  _{\alpha\beta}\tilde{Z}_{ab}-2\left(  \gamma^{a}C\right)
_{\alpha\beta}P_{a}\right]  ,\\
\left\{  Q_{\alpha},\Sigma_{\beta}\right\}   &  =-\frac{1}{2}\left(
\gamma^{ab}C\right)  _{\alpha\beta}Z_{ab},\\
\left[  J_{ab},\tilde{Z}_{cd}\right]   &  =\eta_{bc}\tilde{Z}_{ad}+\eta
_{ad}\tilde{Z}_{bc}-\eta_{bd}\tilde{Z}_{ac}-\eta_{ac}\tilde{Z}_{bd},\\
\left[  \tilde{Z}_{ab},\tilde{Z}_{cd}\right]   &  =\eta_{bc}Z_{ad}+\eta
_{ad}Z_{bc}-\eta_{bd}Z_{ac}-\eta_{ac}Z_{bd},\\
\text{others}  &  =0.
\end{align*}
This algebra can be found by an $\mathrm{S}$-expansion of $\mathfrak{osp}%
(4/1)$ superalgebra.\textbf{~}\cite{eve1,azc}.

In order to write down a four dimensional ChSAS\textbf{\ }action, we start by
expressing the gauge fields $A$ and $B$ at the base of Maxwell superalgebra.

To interpret the gauge field associated with a traslational generator $P_{a}$
as the vielbein, one is \textsl{forced} to introduce a length scale $\ell$ in
the theory. Since one can always chooses Lie algebra generators $T_{A}$ to be
dimensionless as well, the one-form connection fields $A=A_{\mu}^{A}%
T_{A}\mathrm{d}x^{\mu}$ must also be dimensionless. However, the vielbein
$e^{a}=e_{\text{ \ }\mu}^{a}\mathrm{d}x^{\mu}$ must have dimensions of length
if it is related to the spacetime metric $g_{\mu\nu}$ through the usual
equation $g_{\mu\nu}=e_{\text{ \ }\mu}^{a}e_{\text{ \ }\nu}^{b}\eta_{ab}.$
This means that the \textquotedblleft true\textquotedblright\ gauge field must
be of the form $e^{a}/\ell$, with $\ell$ a length parameter. Therefore,
following Refs.~\cite{Cha89}, \cite{Cha90}, the one-form gauge field $A$ is
given by%
\[
A=\frac{1}{\ell}e^{a}P_{a}+\frac{1}{2}\omega^{ab}J_{ab}+\frac{1}{2}%
k^{ab}Z_{ab}+\frac{1}{2}\tilde{k}^{ab}\tilde{Z}_{ab}+\frac{1}{\sqrt{\ell}}%
\psi^{\alpha}Q_{\alpha}+\frac{1}{\sqrt{\ell}}\xi^{\alpha}\Sigma_{\alpha},
\]
where $e^{a}$ is identified as the $1$\textbf{-}form vierbein, $\omega^{ab}$
is the $1$\textbf{-}form spin\textbf{ }connection, $k^{ab}$and $\tilde{k}%
^{ab}$ are extra antisymmetric bosonic $1$-form fields, and $\psi^{\alpha}$,
$\xi^{\alpha}$ are fermionic $1$-form fields. The corresponding $2$-form
curvature is given by%
\[
F=\frac{1}{\ell}\hat{T}^{a}P_{a}+\frac{1}{2}R^{ab}J_{ab}+\frac{1}{2}%
f^{ab}Z_{ab}+\frac{1}{2}\tilde{f}^{ab}\tilde{Z}_{ab}+\frac{1}{\sqrt{\ell}}%
\Psi^{\alpha}Q_{\alpha}+\frac{1}{\sqrt{\ell}}\Xi^{\beta}\Sigma_{\beta},
\]
with%
\begin{align*}
\hat{T}^{a}  &  =T^{a}+\frac{1}{2}\bar{\psi}\gamma^{a}\psi,\\
R^{ab}  &  =\mathrm{d}\omega^{ab}+\omega_{\text{ \ }c}^{a}\omega^{cb},\\
f^{ab}  &  =\mathrm{D}k^{ab}+\frac{1}{\ell^{2}}e^{a}e^{b}+\tilde{k}_{\text{
\ }c}^{a}\tilde{k}^{cb}-\frac{1}{\ell}\bar{\xi}\gamma^{ab}\psi,\\
\tilde{f}^{ab}  &  =\mathrm{D}\tilde{k}^{ab}-\frac{1}{2\ell}\bar{\psi}%
\gamma^{ab}\psi,\\
\Psi^{\alpha}  &  =\mathrm{D}\psi^{\alpha},\\
\Xi^{\beta}  &  =\mathrm{D}\xi^{\beta}-\frac{1}{4}\tilde{k}^{ab}\psi^{\alpha
}\left(  \gamma_{ab}\right)  _{\alpha}^{\text{ \ }\beta}-\frac{1}{2\ell}%
e^{a}\psi^{\alpha}\left(  \gamma_{a}\right)  _{\alpha}^{\text{ \ }\beta}.
\end{align*}
For the $2$-form $B$, we can write%
\[
B=B^{a}P_{a}+\frac{1}{2}B^{ab}J_{ab}+\frac{1}{2}\beta^{ab}Z_{ab}+\frac{1}%
{2}\tilde{\beta}^{ab}\tilde{Z}_{ab}+\lambda^{\alpha}Q_{\alpha}+\chi^{\alpha
}\Sigma_{\alpha},
\]
where $B^{a},\;B^{ab},\;\beta^{ab},$ $\tilde{\beta}^{ab},$ $\lambda^{\alpha},$
$\chi^{\alpha}$ are $2$-forms that we must determine. The
corresponding\textbf{\ }$3$-form curvature is given by%
\[
H=H^{a}P_{a}+\frac{1}{2}H^{ab}J_{ab}+\frac{1}{2}\Theta^{ab}Z_{ab}+\frac{1}%
{2}\tilde{\Theta}^{ab}\tilde{Z}_{ab}+\tilde{H}^{\alpha}Q_{\alpha}%
+\mathcal{H}^{\alpha}\Sigma_{\alpha},
\]
with,%
\begin{align}
H^{a}  &  =\mathrm{D}B^{a}-\frac{1}{\ell}B_{\text{ \ }c}^{a}e^{c}+\frac
{1}{\sqrt{\ell}}\bar{\psi}\gamma^{a}\lambda,\label{fda1}\\
H^{ab}  &  =\mathrm{D}B^{ab},\label{fda2}\\
\Theta^{ab}  &  =\mathrm{D}\beta^{ab}-B_{\text{ \ }c}^{[a|}k^{c|b]}-\frac
{1}{\ell}B^{[a|}e^{|b]}-\tilde{\beta}_{\text{ \ }c}^{[a|}\tilde{k}%
^{c|b]}-\frac{1}{\sqrt{\ell}}\bar{\psi}\gamma^{ab}\chi+\frac{1}{\sqrt{\ell}%
}\bar{\lambda}\gamma^{ab}\xi,\nonumber\\
\tilde{\Theta}^{ab}  &  =\mathrm{D}\tilde{\beta}^{ab}-B_{\text{ \ }c}%
^{[a|}\tilde{k}^{c|b]}-\frac{1}{\sqrt{\ell}}\bar{\psi}\gamma^{ab}%
\lambda,\label{fda4}\\
\tilde{H}^{\alpha}  &  =\mathrm{D}\lambda^{\alpha}+\frac{1}{4\sqrt{\ell}%
}B^{ab}\psi^{\beta}\left(  \gamma_{ab}\right)  _{\beta}^{\text{ \ }\alpha
},\label{fda5}\\
\mathcal{H}^{\alpha}  &  =\mathrm{D}\chi^{\alpha}+\frac{1}{4\sqrt{\ell}}%
B^{ab}\xi^{\beta}\left(  \gamma_{ab}\right)  _{\beta}^{\text{ \ }\alpha}%
-\frac{1}{2\ell}e^{a}\lambda^{\beta}\left(  \gamma_{a}\right)  _{\beta
}^{\text{ \ }\alpha}+\frac{1}{2\sqrt{\ell}}B^{a}\psi^{\beta}\left(  \gamma
_{a}\right)  _{\beta}^{\text{ \ }\alpha}\nonumber\\
&  -\frac{1}{4}\tilde{k}^{ab}\lambda^{\beta}\left(  \gamma_{ab}\right)
_{\beta}^{\text{ \ }\alpha}+\frac{1}{4\sqrt{\ell}}\tilde{\beta}^{ab}%
\psi^{\beta}\left(  \gamma_{ab}\right)  _{\beta}^{\text{ \ }\alpha}.
\label{fda6}%
\end{align}
The problem now is to express the form $B$\ defined by the equations
~(\ref{fda1}-\ref{fda6}) in terms of the one-forms $\{e^{a},\omega^{ab}%
,k^{ab},\tilde{k}^{ab},\psi^{\alpha},\xi^{\alpha}\}$ of the Maxwell
superalgebra. To express the $2$-forms $\left\{  B^{a},B^{ab},\beta
^{ab},\tilde{\beta}^{ab},\lambda,\chi\right\}  $ as the wedge product of the
$1$-forms $\{e^{a},\omega^{ab},k^{ab},\tilde{k}^{ab},\psi^{\alpha},\xi
^{\alpha}\}$ we follow a procedure developed in Refs.~\cite{dauria,azcar}.
Imposing the ansatz%
\begin{align}
B^{a}  &  =\frac{a_{1}}{2\ell}\omega_{\text{ \ }b}^{a}e^{b}+\frac{a_{2}}%
{2\ell}k_{\text{ \ }b}^{a}e^{b}+\frac{a_{3}}{2\ell}\tilde{k}_{\text{ \ }b}%
^{a}e^{b}+\frac{a_{4}}{\ell}\bar{\psi}\gamma^{a}\psi+\frac{a_{5}}{\ell}%
\bar{\psi}\gamma^{a}\xi+\frac{a_{6}}{\ell}\bar{\xi}\gamma^{a}\xi,\label{b1}\\
B^{ab}  &  =\frac{b_{1}}{2\ell^{2}}e^{a}e^{b}+\frac{b_{2}}{2}\omega_{\text{
\ }c}^{[a|}k^{cb]}+\frac{b_{3}}{2}k_{\text{ \ }c}^{a}k^{cb}+\frac{b_{4}}%
{2}\omega_{\text{ \ }c}^{[a|}\tilde{k}^{c|b]}+\frac{b_{5}}{2}\tilde{k}_{\text{
\ }c}^{a}\tilde{k}^{cb}\nonumber\\
&  +\frac{b_{6}}{2}\omega_{\text{ \ }c}^{a}\omega^{cb}+\frac{b_{7}}{\ell}%
\bar{\psi}\gamma^{ab}\psi+\frac{b_{8}}{\ell}\bar{\psi}\gamma^{ab}\xi
+\frac{b_{9}}{\ell}\bar{\xi}\gamma^{ab}\xi,\label{b2}\\
\beta^{ab}  &  =\frac{c_{1}}{2\ell^{2}}e^{a}e^{b}+\frac{c_{2}}{2}%
\omega_{\text{ \ }c}^{[a|}k^{c|b]}+\frac{c_{3}}{2}k_{\text{ \ }c}^{a}%
k^{cb}+\frac{c_{4}}{2}\omega_{\text{ \ }c}^{[a|}\tilde{k}^{c|b]}+\frac{c_{5}%
}{2}\tilde{k}_{\text{ \ }c}^{a}\tilde{k}^{cb}\nonumber\\
&  +\frac{c_{6}}{2}\omega_{\text{ \ }c}^{a}\omega^{cb}+\frac{c_{7}}{\ell}%
\bar{\psi}\gamma^{ab}\psi+\frac{c_{8}}{\ell}\bar{\psi}\gamma^{ab}\xi
+\frac{c_{9}}{\ell}\bar{\xi}\gamma^{ab}\xi,\label{b3}\\
\tilde{\beta}^{ab}  &  =\frac{d_{1}}{2\ell^{2}}e^{a}e^{b}+\frac{d_{2}}%
{2}\omega_{\text{ \ }c}^{[a|}k^{c|b]}+\frac{d_{3}}{2}k_{\text{ \ }c}^{a}%
k^{cb}+\frac{d_{4}}{2}\omega_{\text{ \ }c}^{[a|}\tilde{k}^{c|b]}+\frac{d_{5}%
}{2}\tilde{k}_{\text{ \ }c}^{a}\tilde{k}^{cb}\nonumber\\
&  +\frac{d_{6}}{2}\omega_{\text{ \ }c}^{a}\omega^{cb}+\frac{d_{7}}{\ell}%
\bar{\psi}\gamma^{ab}\psi+\frac{d_{8}}{\ell}\bar{\psi}\gamma^{ab}\xi
+\frac{d_{9}}{\ell}\bar{\xi}\gamma^{ab}\xi,\label{b4}\\
\lambda_{\alpha}  &  =\frac{f_{1}}{\ell}e_{a}\gamma^{a}\psi_{\alpha}%
+\frac{f_{2}}{\ell}e_{a}\gamma^{a}\xi_{\alpha}+\frac{f_{3}}{2}\omega
_{ab}\gamma^{ab}\psi_{\alpha}+\frac{f_{4}}{2}\omega_{ab}\gamma^{ab}\xi
_{\alpha}\nonumber\\
&  +\frac{f_{5}}{2}k_{ab}\gamma^{ab}\psi_{\alpha}+\frac{f_{6}}{2}k_{ab}%
\gamma^{ab}\xi_{\alpha}+\frac{f_{7}}{2}\tilde{k}_{ab}\gamma^{ab}\psi_{\alpha
}+\frac{f_{8}}{2}\tilde{k}_{ab}\gamma^{ab}\xi_{\alpha},\label{b5}\\
\chi_{\alpha}  &  =\frac{g_{1}}{\ell}e_{a}\gamma^{a}\psi_{\alpha}+\frac{g_{2}%
}{\ell}e_{a}\gamma^{a}\xi_{\alpha}+\frac{g_{3}}{2}\omega_{ab}\gamma^{ab}%
\psi_{\alpha}+\frac{g_{4}}{2}\omega_{ab}\gamma^{ab}\xi_{\alpha}\nonumber\\
&  +\frac{g_{5}}{2}k_{ab}\gamma^{ab}\psi_{\alpha}+\frac{g_{6}}{2}k_{ab}%
\gamma^{ab}\xi_{\alpha}+\frac{g_{7}}{2}\tilde{k}_{ab}\gamma^{ab}\psi_{\alpha
}+\frac{g_{8}}{2}\tilde{k}_{ab}\gamma^{ab}\xi_{\alpha}, \label{b6}%
\end{align}
where $a_{1},\ldots,a_{6},b_{1},\ldots,b_{9},c_{1},\ldots,c_{9},d_{1}%
,\ldots,d_{9},$ $f_{1},\ldots,f_{8},g_{1},\ldots,g_{8}$ are arbitrary
constants, and introducing eqs.~(\ref{b1}-\ref{b6}) in eqs.~(\ref{fda1}%
-\ref{fda6}), when $H^{a}=H^{ab}=\Theta^{ab}=\tilde{\Theta}^{ab}=\tilde
{H}^{\alpha}=\mathcal{H}^{\alpha}=0$, we find%
\begin{align}
B^{a}  &  =\frac{a_{4}}{\ell}\bar{\psi}\gamma^{a}\psi+\frac{a_{5}}{\ell}%
\bar{\psi}\gamma^{a}\xi,\label{e1}\\
B^{ab}  &  =0,\label{e2}\\
\beta^{ab}  &  =\frac{c_{1}}{2\ell^{2}}e^{a}e^{b}+\frac{c_{5}}{2}\tilde
{k}_{\text{ \ }c}^{a}\tilde{k}^{cb}+\frac{c_{8}}{\ell}\bar{\psi}\gamma^{ab}%
\xi+\frac{c_{9}}{\ell}\bar{\xi}\gamma^{ab}\xi,\label{e3}\\
\tilde{\beta}^{ab}  &  =\frac{d_{7}}{\ell}\bar{\psi}\gamma^{ab}\psi
+\frac{d_{8}}{\ell}\bar{\psi}\gamma^{ab}\xi,\label{e4}\\
\lambda_{\alpha}  &  =\frac{f_{1}}{\ell}e_{a}\gamma^{a}\psi_{\alpha}%
+\frac{f_{7}}{2}\tilde{k}_{ab}\gamma^{ab}\psi_{\alpha},\label{e5}\\
\chi_{\alpha}  &  =\frac{g_{1}}{\ell}e_{a}\gamma^{a}\psi_{\alpha}+\frac{g_{2}%
}{\ell}e_{a}\gamma^{a}\xi_{\alpha}+\frac{g_{7}}{2}\tilde{k}_{ab}\gamma
^{ab}\psi_{\alpha}+\frac{g_{8}}{2}\tilde{k}_{ab}\gamma^{ab}\xi_{\alpha}.
\label{e6}%
\end{align}
The fields given by eqs.~\textbf{(}\ref{e1}\textbf{-}\ref{e6}\textbf{)}
represent the most general solution that can be built from the fields
$\{e^{a},\omega^{ab},k^{ab},\tilde{k}^{ab},\psi^{\alpha},\xi^{\alpha}\}$. Any
choice of the constants represent a solution to the FDA.

\subsection{Chern-Simons-Antoniadis-Savvidy form}

Using the invariant tensor found in Ref. \cite{eve1}%
\begin{align*}
\left\langle J_{ab}J_{cd}\right\rangle  &  =\alpha_{0}\epsilon_{abcd},\text{
\ \ }\left\langle J_{ab}\tilde{Z}_{cd}\right\rangle =\alpha_{2}\epsilon
_{abcd},\\
\left\langle \tilde{Z}_{ab}\tilde{Z}_{cd}\right\rangle  &  =\alpha_{4}%
\epsilon_{abcd},\text{ \ \ }\left\langle J_{ab}Z_{cd}\right\rangle =\alpha
_{4}\epsilon_{abcd},\\
\left\langle Q_{\alpha}Q_{\beta}\right\rangle  &  =2\alpha_{2}\left(
\gamma_{5}\right)  _{\alpha\beta},\text{ \ \ }\left\langle Q_{\alpha}%
\Sigma_{\beta}\right\rangle =2\alpha_{4}\left(  \gamma_{5}\right)
_{\alpha\beta},
\end{align*}
being $\alpha_{0}$, $\alpha_{2}$ and $\alpha_{4}$ dimensionless arbitrary
independent constants, the Chern--Simons--Antoniadis--Savvidy Lagrangian
$\mathcal{L}_{\mathrm{ChSAS}}^{(4)}\equiv\mathfrak{C}_{\mathrm{ChSAS}}^{(4)}$
is explicitly given by%
\begin{align}
\mathcal{L}_{\mathrm{ChSAS}}^{(4)}  &  =\frac{1}{4}\epsilon_{abcd}\left(
\alpha_{0}R^{ab}B^{cd}+\alpha_{4}\left(  R^{ab}\beta^{cd}+f^{ab}B^{cd}\right)
\right) \nonumber\\
&  +\frac{1}{4}\epsilon_{abcd}\alpha_{2}\left(  R^{ab}\tilde{\beta}^{cd}%
+\frac{1}{4}\tilde{f}^{ab}B^{cd}\right)  +\alpha_{4}\tilde{f}^{ab}\tilde
{\beta}^{cd}\nonumber\\
&  +\frac{2\alpha_{2}}{\sqrt{\ell}}\Psi^{\alpha}\left(  \gamma_{5}\right)
_{\alpha}^{\text{ \ }\beta}\lambda_{\beta}+\frac{2\alpha_{4}}{\sqrt{\ell}}%
\Psi^{\alpha}\left(  \gamma_{5}\right)  _{\alpha}^{\text{ \ }\beta}\chi
_{\beta}+\frac{2\alpha_{4}}{\sqrt{\ell}}\lambda^{\alpha}\left(  \gamma
_{5}\right)  _{\alpha}^{\text{ \ }\beta}\Xi_{\beta}. \label{ChSAS1}%
\end{align}
Using the FDA expansion given by eqs.\textbf{~(}\ref{e1}\textbf{-}%
\ref{e6}\textbf{), }the\textbf{\ }Chern--Simons--Antoniadis--Savvidy
Lagrangian for the Maxwell algebra takes the form%
\begin{align}
\mathcal{L}_{\mathrm{ChSAS}}^{(4)}  &  =\frac{1}{4}\epsilon_{abcd}\left(
\alpha_{4}R^{ab}\left(  \frac{c_{1}}{2\ell^{2}}e^{c}e^{d}+\frac{c_{5}}%
{2}\tilde{k}_{\text{ \ }f}^{c}\tilde{k}^{fd}+\frac{c_{8}}{\ell}\bar{\psi
}\gamma^{cd}\xi+\frac{c_{9}}{\ell}\bar{\xi}\gamma^{cd}\xi\right)  \right)
\nonumber\\
&  +\frac{1}{4}\epsilon_{abcd}\alpha_{2}R^{ab}\left(  \frac{d_{7}}{\ell}%
\bar{\psi}\gamma^{cd}\psi+\frac{d_{8}}{\ell}\bar{\psi}\gamma^{cd}\xi\right)
+\alpha_{4}\tilde{f}^{ab}\left(  \frac{d_{7}}{\ell}\bar{\psi}\gamma^{cd}%
\psi+\frac{d_{8}}{\ell}\bar{\psi}\gamma^{cd}\xi\right) \nonumber\\
&  +\frac{2}{\ell^{3/2}}\left(  \alpha_{2}f_{1}+\alpha_{4}g_{1}\right)
\Psi^{\alpha}e_{a}\left(  \gamma_{5}\right)  _{\alpha}^{\text{ \ }\beta
}\left(  \gamma^{a}\right)  _{\beta}^{\text{ \ }\gamma}\psi_{\gamma
}\nonumber\\
&  +\left(  \frac{f_{7}}{2}\frac{2\alpha_{2}}{\sqrt{\ell}}+\frac{g_{7}}%
{2}\frac{2\alpha_{4}}{\sqrt{\ell}}\right)  \Psi^{\alpha}\tilde{k}_{ab}\left(
\gamma_{5}\right)  _{\alpha}^{\text{ \ }\beta}\left(  \gamma^{ab}\right)
_{\beta}^{\text{ \ }\gamma}\psi_{\gamma}\nonumber\\
&  +\frac{g_{2}}{\ell}\frac{2\alpha_{4}}{\sqrt{\ell}}\Psi^{\alpha}\left(
\gamma_{5}\right)  _{\alpha}^{\text{ \ }\beta}e_{a}\left(  \gamma^{a}\right)
_{\beta}^{\text{ \ }\gamma}\xi_{\gamma}+\frac{g_{8}}{2}\frac{2\alpha_{4}%
}{\sqrt{\ell}}\Psi^{\alpha}\left(  \gamma_{5}\right)  _{\alpha}^{\text{
\ }\beta}\tilde{k}_{ab}\left(  \gamma^{ab}\right)  _{\beta}^{\text{ \ }\gamma
}\xi_{\gamma}\nonumber\\
&  -\frac{f_{1}}{\ell}\frac{2\alpha_{4}}{\sqrt{\ell}}e_{a}\psi^{\beta}\left(
\gamma^{a}\right)  _{\beta}^{\text{ \ }\alpha}\left(  \gamma_{5}\right)
_{\alpha}^{\text{ \ }\gamma}\Xi_{\gamma}-\frac{f_{7}}{2}\frac{2\alpha_{4}%
}{\sqrt{\ell}}\tilde{k}_{ab}\psi^{\beta}\left(  \gamma^{ab}\right)  _{\beta
}^{\text{ \ }\alpha}\left(  \gamma_{5}\right)  _{\alpha}^{\text{ \ }\gamma}%
\Xi_{\gamma}. \label{ChSAS2}%
\end{align}
From eq. (\ref{ChSAS2}) we can see that if $c_{9}=d_{8}=f_{1}=f_{7}%
=g_{2}=g_{8}=0$, which are conditions \ consistent with the equations
(\ref{b1}-\ref{b6}) and~(\ref{fda1}-\ref{fda6}), we have that $\mathcal{L}%
_{\mathrm{ChSAS}}^{(4)}$ is given by%
\begin{align}
\mathcal{L}_{\mathrm{ChSAS}}^{(4)}  &  =\frac{\alpha_{4}c_{1}}{8}%
\epsilon_{abcd}R^{ab}e^{c}e^{d}+2\alpha_{4}g_{1}\sqrt{\ell}\Psi\gamma_{5}%
e_{a}\gamma^{a}\psi+\frac{\alpha_{4}c_{5}}{8}l^{2}\epsilon_{abcd}R^{ab}%
\tilde{k}_{\text{ \ }f}^{c}\tilde{k}^{fd}\nonumber\\
&  +\left(  \frac{\alpha_{4}c_{8}}{4}+\frac{\alpha_{2}d_{7}}{4}\right)
l\epsilon_{abcd}R^{ab}\bar{\psi}\gamma^{cd}\xi+\frac{\alpha_{4}d_{7}}%
{4}l\epsilon_{abcd}\tilde{f}^{ab}\bar{\psi}\gamma^{cd}\xi\nonumber\\
&  +\alpha_{4}g_{7}l^{3/2}\Psi\gamma_{5}\tilde{k}_{ab}\gamma^{ab}\psi.
\label{ChSAS3}%
\end{align}

Here it is necessary to notice that:

\textbf{(a)} The first two terms contains the Einstein--Hilbert and the
Rarita-Schwinger terms given by $\epsilon_{abcde}R^{ab}e^{c}e^{d}e^{e}$ \ and
$\Psi\gamma_{5}e_{a}\gamma^{a}\psi$ respectively.

\textbf{(b)} The following terms could be interpreted as non-linear couplings
between the bosonic and fermionic "matter" fields $\tilde{k}^{ab}$, $\xi$, the
Rarita-Schwinger field $\psi$ and the curvature, where the parameter $l$ can
be considered as a kind of coupling constant.

From eq.~(\ref{ChSAS3}), we can see that when\textbf{\ }$l\ll1\mathbf{,}$ the
Chern--Simons--Antoniadis--Savvidy Lagrangian for the Maxwell superalgebra is
given by%
\begin{equation}
\mathcal{L}_{\mathrm{ChSAS}}^{(4)}=\frac{\alpha_{4}c_{1}}{8}\epsilon
_{abcd}R^{ab}e^{c}e^{d}+2\alpha_{4}g_{1}\sqrt{\ell}\Psi\gamma_{5}e_{a}%
\gamma^{a}\psi, \label{ChSAS4}%
\end{equation}
where we can see that the Chern--Simons--Savvidy Lagrangian reproduces, except
for numerical coefficients, the Lagrangian for standard supergravity.

It is perhaps interesting to note that the conmutation relation $[P_{a}%
,P_{b}]=Z_{ab}$ depends on the $Z_{ab}$ generators. The consequences on the
Lagrangian of this non-zero bracket are related to the gauge field $k^{ab}$
associated to $Z_{ab}$. If we not consider the $k^{ab}$ dependence, or if we
take a limit on the theory in which the gauge field $k^{ab}$ effects are not
included, then it is not surprising that the curvature term in the above
Lagrangian looks like the standard gravity Lagrangian.

\section{The Extended Cartan Homotopy Formula\textbf{ in }$\left(
2n+2\right)  $-dimensions}

The Extended Cartan Homotopy Formula \textbf{(}ECHF\textbf{) }reads \cite{mss}%
\begin{equation}
\int_{\partial T_{r+1}}\frac{l_{t}^{p}}{p!}\pi=\int_{T_{r+1}}\frac{l_{t}%
^{p+1}}{\left(  p+1\right)  !}\mathrm{d}\pi+\left(  -1\right)  ^{p+q}%
\mathrm{d}\int_{T_{r+1}}\frac{l_{t}^{p+1}}{\left(  p+1\right)  !}\pi,
\label{cehf}%
\end{equation}
where, in this case, $\pi$ represents a polynomial in the forms $\left\{
A_{t},B_{t},F_{t},H_{t},\mathrm{d}_{t}A_{t},\mathrm{d}_{t}F_{t}\right\}  $
which is also an $m$-form on $M$ and a $q$-form on $T_{r+1}$, with $m\geq p$
and $p+q=r$. The exterior derivatives on $M$ and $T_{r+1}$ are denoted
respectively by d and d$_{t}$. The operator $l_{t}$, called homotopy
derivation, maps differential forms on $M$ and $T_{r+1}$ according to%
\[
l_{t}:\Omega^{a}\left(  M\right)  \times\Omega^{b}\left(  T_{r+1}\right)
\rightarrow\Omega^{a-1}\left(  M\right)  \times\Omega^{b+1}\left(
T_{r+1}\right)  ,
\]
and it satisfies Leibniz's rule as well as $\mathrm{d}$ as $\mathrm{d}_{t}$.
In our case, we will consider the polinomial $\pi=\left\langle F_{t}^{n}%
H_{t}\right\rangle $. This choice has the three following properties: (i)
$\pi$ is $M$-closed, i.e., $\mathrm{d}\pi=0$, (ii) $\pi$ is a 0-form on
$T_{r+1}$, and (iii) $\pi$ is a $\left(  2n+3\right)  $-form on $M$. The
allowed values for $p$ are $p=0,\ldots,2n+3$. The ECHF reduces in this case to%
\begin{equation}
\int_{\partial T_{p+1}}\frac{l_{t}^{p}}{p!}\pi=\left(  -1\right)
^{p+q}\mathrm{d}\int_{T_{p+1}}\frac{l_{t}^{p+1}}{\left(  p+1\right)  !}\pi.
\label{cehfc}%
\end{equation}
Since the three operators $\mathrm{d}$, $\mathrm{d}_{t}$ and $l_{t}$ define a
graded algebra given by \cite{mss}
\begin{align}
\mathrm{d}^{2}  &  =0,\text{ \ \ }\mathrm{d}_{t}^{2}=0,\text{ \ \ }\left\{
\mathrm{d},\mathrm{d}_{t}\right\}  =0,\\
\left[  l_{t},\mathrm{d}\right]   &  =\mathrm{d}_{t},\text{ \ \ }\left[
l_{t},\mathrm{d}_{t}\right]  =0,
\end{align}
we have that the action of $l_{t}$ on $\{A_{t}$, $B_{t}$, $F_{t}$, $H_{t}$,
$\mathrm{d}_{t}A_{t}$, $\mathrm{d}_{t}F_{t}\}$ reads \cite{mss}%
\[
l_{t}A_{t}=0,\text{ }l_{t}F_{t}=l_{t}\left(  \mathrm{d}A_{t}+A_{t}%
A_{t}\right)  =\left(  \mathrm{d}l_{t}+\mathrm{d}_{t}\right)  A_{t}%
=\mathrm{d}_{t}A_{t},
\]
while the action of $l_{t}$ on $B_{t}$ and $H_{t}$ must be determined.

Particular cases of (\ref{cehfc}) with $\pi$ given by (\ref{drei}) which we
review below, reproduce both the Chern--Weil Theorem and the Triangle Formula.
In fact, when $p=0,$ we find that Eq.(\ref{cehfc}) takes the form,%
\begin{equation}
\int_{\partial T_{1}}\pi=\mathrm{d}\int_{T_{1}}l_{t}\pi, \label{p0}%
\end{equation}
where $\pi=\left\langle F_{t}^{n}H_{t}\right\rangle $ and $A_{t}=A_{0}%
+t\Theta,$ $B_{t}=B_{0}+t\Phi$. The left side of (\ref{p0}) is given by%
\[
\int_{\partial T_{1}}\left\langle F_{t}^{n},H_{t}\right\rangle =\int_{0}%
^{1}\mathrm{d}t\left\langle F_{t}^{n},H_{t}\right\rangle =\left\langle
F_{1}^{n},H_{1}\right\rangle -\left\langle F_{0}^{n},H_{0}\right\rangle ,
\]
while for the right side we have%
\[
\mathrm{d}\int_{T_{1}}l_{t}\left\langle F_{t}^{n},H_{t}\right\rangle
=\mathrm{d}\left\{  n\int_{0}^{1}\mathrm{d}t\left\langle F_{t}^{n-1}%
,\Theta,H_{t}\right\rangle +\int_{T_{1}}\left\langle F_{t}^{n},l_{t}%
H_{t}\right\rangle \right\}  ,
\]
so that%
\[
\left\langle F_{1}^{n},H_{1}\right\rangle -\left\langle F_{0}^{n}%
,H_{0}\right\rangle =\mathrm{d}\left\{  n\int_{0}^{1}dt\left\langle
F_{t}^{n-1},\Theta,H_{t}\right\rangle +\int_{T_{1}}\left\langle F_{t}%
^{n},l_{t}H_{t}\right\rangle \right\}  .
\]
From the Chern--Weil theorem and $B_{t}=B_{0}+t\Phi$, we see that%
\[
\left\langle F_{t}^{n},l_{t}H_{t}\right\rangle =\left\langle F_{t}%
^{n},\mathrm{d}_{t}B_{t}\right\rangle ,
\]
so that $l_{t}H_{t}=\mathrm{d}_{t}B_{t}.$\ On the other hand, we have%
\[
\left\langle F_{t}^{n},l_{t}\left(  \mathrm{d}B_{t}+A_{t}B_{t}-B_{t}%
A_{t}\right)  \right\rangle ~=\left\langle F_{t}^{n},\mathrm{d}_{t}%
B_{t}\right\rangle ,
\]%
\[
\left\langle F_{t}^{n},\left(  \left(  \mathrm{d}+A_{t}\right)  \left(
l_{t}B_{t}\right)  +\mathrm{d}_{t}B_{t}-\left(  l_{t}B_{t}\right)
A_{t}\right)  \right\rangle =\left\langle F_{t}^{n},\mathrm{d}_{t}%
B_{t}\right\rangle ,
\]
and therefore $\left\langle F_{t}^{n},\mathrm{D}_{t}l_{t}B_{t}\right\rangle
=\mathrm{d}\left\langle F_{t}^{n},l_{t}B_{t}\right\rangle =0$ and $l_{t}%
B_{t}=0.$ Summarizing, we can write $l_{t}B_{t}=0$ and $l_{t}H_{t}%
=\mathrm{d}_{t}B_{t}.$

\subsection{The subspace separation method}

In this subsection we will show that the subspace separation method developed
in Ref. \cite{irs} can be generalized to the case of
Chern--Simons--Antoniadis--Savvidy formalism. This means that the so called
triangle equation (\ref{p7}) splits the transgression form $\mathfrak{T}%
^{\left(  2n+2\right)  }\left(  A_{2},B_{2};A_{0},B_{0}\right)  $ into the sum
of two transgressions forms depending on an intermediate connection
$A_{1},B_{1}$ plus a exact form \ $\mathfrak{T}^{\left(  2n+1\right)  }\left(
A_{2},B_{2};A_{1},B_{1};A_{0},B_{0}\right)  $ shown in Eq.\ (\ref{p5}).

When $p=1$ we have that Eq.~(\ref{cehfc}) is given by%
\begin{equation}
\int_{\partial T_{2}}l_{t}\left\langle F_{t}^{n},H_{t}\right\rangle
=-\mathrm{d}\int_{T_{2}}\frac{l_{t}^{2}}{2}\left\langle F_{t}^{n}%
,H_{t}\right\rangle , \label{p2}%
\end{equation}
where%
\begin{align*}
A_{t}  &  =t^{0}\left(  A_{0}-A_{1}\right)  +t^{2}\left(  A_{2}-A_{1}\right)
+A_{1},\\
B_{t}  &  =t^{0}\left(  B_{0}-B_{1}\right)  +t^{2}\left(  B_{2}-B_{1}\right)
+B_{1}.
\end{align*}
The left side of (\ref{p2}) corresponds to an integral along the boundary of
the simplex $T_{2}=\left(  A_{2},B_{2};A_{1},B_{1};A_{0},B_{0}\right)  :$%
\begin{align}
\int_{\partial T_{2}}l_{t}\left\langle F_{t}^{n},H_{t}\right\rangle  &
=\mathfrak{T}^{\left(  2n+2\right)  }\left(  A_{2},B_{2};A_{1},B_{1}\right)
-\mathfrak{T}^{\left(  2n+2\right)  }\left(  A_{2},B_{2};A_{0},B_{0}\right)
\nonumber\\
&  +\mathfrak{T}^{\left(  2n+2\right)  }\left(  A_{1},B_{1};A_{0}%
,B_{0}\right)  . \label{p3}%
\end{align}
The right side of (\ref{p2}) is given by%
\begin{equation}
\mathrm{d}\int_{T_{2}}\frac{l_{t}^{2}}{2}\left\langle F_{t}^{n},H_{t}%
\right\rangle =\mathrm{d}\mathfrak{T}^{\left(  2n+1\right)  }\left(
A_{2},B_{2};A_{1},B_{1};A_{0},B_{0}\right)  , \label{p4}%
\end{equation}
where%
\begin{align}
\mathfrak{T}^{\left(  2n+1\right)  }\left(  A_{2},B_{2};A_{1},B_{1}%
;A_{0},B_{0}\right)   &  =\int_{0}^{1}\mathrm{d}t\int_{0}^{t}\mathrm{d}%
s\left\{  n\left(  n-1\right)  \left\langle F_{t}^{n-1},\left(  A_{2}%
-A_{1}\right)  ,\left(  A_{1}-A_{0}\right)  ,H_{t}\right\rangle \right.
\nonumber\\
&  +n\left\langle F_{t}^{n-1},A_{0},\left(  B_{2}-B_{1}\right)  \right\rangle
+n\left\langle F_{t}^{n-1},A_{1},\left(  B_{0}-B_{2}\right)  \right\rangle \\
&  \left.  +n\left\langle F_{t}^{n-1},A_{2},\left(  B_{1}-B_{0}\right)
\right\rangle \right\}  . \label{p5}%
\end{align}
In (\ref{p5}) we have introduced dummy parameters $t=1-t^{0}$ and $s=t^{2}$,
in terms of which $A_{t}$ reads%
\begin{equation}
A_{t}=A_{0}+t(A_{1}-A_{0})+s(A_{2}-A_{1}).
\end{equation}
Thus we have that the triangle equation is given by%
\begin{align}
&  \mathfrak{T}^{\left(  2n+2\right)  }\left(  A_{2},B_{2};A_{1},B_{1}\right)
-\mathfrak{T}^{\left(  2n+2\right)  }\left(  A_{2},B_{2};A_{0},B_{0}\right)
+\mathfrak{T}^{\left(  2n+2\right)  }\left(  A_{1},B_{1};A_{0},B_{0}\right)
\nonumber\\
&  =-\mathrm{d}\mathfrak{T}^{\left(  2n+1\right)  }\left(  A_{2},B_{2}%
;A_{1},B_{1};A_{0},B_{0}\right)  , \label{p6}%
\end{align}
or alternatively%
\begin{align}
\mathfrak{T}^{\left(  2n+2\right)  }\left(  A_{2},B_{2};A_{0},B_{0}\right)  =
&  \mathfrak{T}^{\left(  2n+2\right)  }\left(  A_{2},B_{2};A_{1},B_{1}\right)
+\mathfrak{T}^{\left(  2n+2\right)  }\left(  A_{1},B_{1};A_{0},B_{0}\right)
\nonumber\\
&  +\mathrm{d}\mathfrak{T}^{\left(  2n+1\right)  }\left(  A_{2},B_{2}%
;A_{1},B_{1};A_{0},B_{0}\right)  . \label{p7}%
\end{align}
We would like to stress that use of the Extended Cartan Homotopy Formula has
allowed us to pinpoint the exact form of the boundary contribution
\ $\mathfrak{T}^{\left(  2n+1\right)  }\left(  A_{2},B_{2};A_{1},B_{1}%
;A_{0},B_{0}\right)  .$

Note that if we choose $A_{0}=0$ and $B_{0}=0$ we obtain an expression that
relates the form Antoniadi--Savvidy transgression form to two
Chern-Simons-Savvidy forms and a total derivative%
\begin{equation}
\mathfrak{T}^{\left(  2n+2\right)  }\left(  A_{2},B_{2};A_{1},B_{1}\right)
=\mathfrak{C}_{\mathrm{ChSAS}}^{\left(  2n+2\right)  }\left(  A_{2}%
,B_{2}\right)  -\mathfrak{C}_{\mathrm{ChSAS}}^{\left(  2n+2\right)  }\left(
A_{1},B_{1}\right)  -\mathrm{d}\mathfrak{T}^{\left(  2n+1\right)  }\left(
A_{2},B_{2};A_{1},B_{1};0,0\right)  .
\end{equation}

\section{Concluding Remarks}

In this Letter some features of the $\left(  2n+2\right)  $-dimensional
transgressions and Chern--Simons--Antoniadis--Savvidy forms used as
Lagrangians for supergravity theories were briefly reviewed. The 2--form field
$B$ can be discomposed in terms of components of the 1-form $A$. It is
performed in a self--consistent way by considering the generalization of
Maurer--Cartan approach to forms of higher order, i.e., free differential
algebras, and by following the procedure used in Refs.
\cite{dauria,azcar,salg,ims}. The final result is a four-dimensional
supergravity action, which is gauge quasi--invariant under the Maxwell
superalgebra. These 4-dimensional results shown that an interesting problem is
to extract physical information from the $\left(  2n+2\right)  $-dimensional
Lagrangian (\ref{CSS general}). A crucial step in this direction is the
separation of the Lagrangian in a way that reflects the inner subspace
structure of the gauge algebra. This is specially interesting in the case of
higher-dimensional supergravity, where superalgebras come naturally split into
distinct subspaces. \ Examples of the use of the method within the
transgression/Chern--Simons--Antoniadis--Savvidy framework will be studied elsewhere.

\begin{acknowledgement}
This work was supported in part by FONDECYT grants 1130653 and 1150719 from
the Government of Chile. One of the authors (SS) was suported by grants
21140490 from CONICYT (National Commission for Scientific and Technological
Re-search, spanish initials) and from Universidad de Concepci\'{o}n, Chile.
\end{acknowledgement}

\end{document}